\documentclass[12pt,a4paper]{article}
\usepackage{amsmath,amssymb,mathrsfs,framed}
\usepackage[colorlinks]{hyperref}
\usepackage{color}
\usepackage{graphicx}

\newcommand{\curl}{{\operatorname{curl}}}

\newcommand{\rem}[1]{}
\newcommand{\bc}{{\mathbf{c}}}

\newcommand{\de}{{\rm d}}

\newcommand{\z}{{\mathbf{z}}}
\newcommand{\bq}{{\mathbf{x}}}
\newcommand{\bv}{{\mathbf{v}}}

\newcommand{\bk}{{\mathbf{k}}}

\newcommand{\tP}{{\widetilde{\Bbb{P}}_1}}
\newcommand{\bP}{{\widetilde{\Bbb{P}}_e}}

\newcommand{\bx}{{\mathbf{x}}}

\newcommand{\bA}{{\mathbf{A}}}
\newcommand{\tA}{{\,\widetilde{\!\boldsymbol{A}\,}_{\!1}}}

\newcommand{\bK}{{\mathbf{K}}}
\newcommand{\bE}{{\mathbf{E}}}
\newcommand{\bB}{{\mathbf{B}}}
\newcommand{\bJ}{{\mathbf{J}}}
\newcommand{\tB}{{\widetilde{\mathbf{B}}_1}}

\newcommand{\bV}{{\boldsymbol{V}}}

\newcommand{\tV}{{\,\widetilde{\boldsymbol{V}}_{\!1}}}

\newcommand{\bmu}{\bar{\mu}}

\newcommand{\ben}{\begin{eqnarray}}
\newcommand{\een}{\end{eqnarray}}

\begin{document}

\title{Electron inertia and quasi-neutrality\\in the Weibel instability}
\author{Enrico Camporeale$^1$, Cesare Tronci$^2$\\
\it\footnotesize $^1$Center for Mathematics and Computer Science (CWI), 1098 XG Amsterdam, The Netherlands\\
\it\footnotesize $^2$Department of Mathematics, University of Surrey, Guildford GU2 7XH, United Kingdom}
\date{\sf\small  Special issue contribution, on the occasion of the Vlasovia 2016 conference}
\maketitle

\begin{abstract}\footnotesize
While electron kinetic effects are well known to be of fundamental importance in several situations, the electron mean-flow inertia is often neglected when lengthscales below the electron skin depth become irrelevant. This has led to the formulation of different reduced models, where electron inertia terms are discarded while retaining some or all kinetic effects. Upon considering general full-orbit particle trajectories, this paper compares the dispersion relations emerging from such models in the case of the Weibel instability. As a result, the question of how lengthscales below the electron skin depth can be neglected in a kinetic treatment emerges as an unsolved problem, since all current theories suffer from drawbacks of different nature. Alternatively, we discuss fully kinetic theories that remove all these drawbacks by restricting to frequencies well below the plasma frequency of both ions and electrons. By giving up on the lengthscale restrictions appearing in previous works, these  models are obtained by assuming quasi-neutrality in the full Maxwell-Vlasov system. 

\end{abstract}

{\footnotesize
\tableofcontents
}

\section{Introduction: Ohm's law and electron inertia}

Electron kinetic effects play a crucial role in a variety of situations. For example, the development of non-gyrotropic components in the electron pressure tensor is a well-known mechanism that drives collisionless magnetic reconnection (see, e.g., \cite{camporeale05, aunai2013,haynes,cazzola2016,swisdak16}). Indeed, the non-gyrotropic electron pressure is among the main mechanisms driving fast reconnection at lengthscales bigger than the plasma skin depth (also known as \emph{electron inertial length}). More specifically, collisionless reconnection is produced by the last two (non ideal) terms in the electron momentum equation
\begin{equation}\label{ohmslaw}
q_e n_e \bE=q_i n_i\bV_{\!i}\times\bB-\bJ\times\bB+\nabla\cdot\bP+m_e n_e\frac{D\bV_{\!e}}{Dt}
\,,
\end{equation}
with the definitions
\begin{align*}
&n_k(\bx,t)=\int f_k(\bx,\bv,t)\,\de\bv
\,,\\
&\bV_{\!k}(\bx,t)=n_k^{-1}\int f_k(\bx,\bv,t)\,\de\bv
\,,\\
&\widetilde{\Bbb{P}}_k(\bx,t)=m_k\int(\bv-\bV_{\!k})(\bv-\bV_{\!k})f_k(\bx,\bv,t)\,\de\bv
\,.
\end{align*}
Here,  $f_k(\bx,\bv,t)$ is the phase-space density of the $k-$th particle species and $D/Dt=\partial/\partial t+\bV_{\!e}\cdot\nabla$ is the convective derivative.
Upon neglecting the displacement current (so that $\bJ=\mu_0^{-1}\nabla\times\bB$) and by invoking quasi-neutrality (so that $n_i=-q_en_e/q_i$), one obtains the generalized Ohm's law in the form
\begin{equation}\label{ohmslaw1}
 \bE=-\bV_{\!i}\times\bB-\frac1{\mu_0e n_e}\,\bB\times(\nabla\times\bB)-\frac1{e n_e}\nabla\cdot\bP-\frac{m_e}e\frac{D\bV_{\!e}}{Dt}
 \,,
\end{equation}
where we have used the notation $q_i=Ze=-Zq_e$ . Each term on the right hand side of Ohm's law has been extensively studied in terms of its contribution to the reconnection flux \cite{cai97,wang00,GEM}. The last term is associated to the inertia of the electron mean-flow and this generates microscopic instabilities at the scale of the skin depth $\delta_e=c/\omega_{pe}$, which can then drive reconnection. However, these lengthscales are often neglected in reduced reconnection models by discarding the electron mean-flow inertia term, so that Ohm's law becomes
\begin{equation}\label{ohmslaw2}
 \bE=-\bV_{\!i}\times\bB-\frac1{\mu_0e n_e}\,\bB\times(\nabla\times\bB)-\frac1{e n_e}\nabla\cdot\bP
 \,,
\end{equation}
This reduced form of Ohm's law has been adopted in a variety of works  \cite{HeWi94b,HeWi94c,KuHeBi,KuHeWi98,KuHeWi,YiWi,YiWiGaBi,WiHe}. In these works, equation  \eqref{ohmslaw2} is combined with a moment truncation for the electron pressure dynamics, which is then coupled to ion motion in either fluid or kinetic description.

\cite{CJ} followed a different strategy for obtaining a reduced model. While retaining small lengthscales, their approach neglected high frequencies by adopting the quasi-neutral limit of the Maxwell-Vlasov system. More specifically, using Amp\`ere's law leads to rewriting { (with no approximation)} the generalized Ohm's law \eqref{ohmslaw1}  as
%\begin{multline*}
%e^2n_e\left(\frac{1}{m_e}+\frac{Z}{m_i}\right)\bE+\mu_0^{-1}\nabla\times\nabla\times\bE
%=-
%e^2n_e\left(\frac{1}{m_e}+\frac{Z}{m_i}\right)\bV_{\!i}\times\bB-\frac{e}{\mu_0m_e}\bB\times(\nabla\times\bB)
%\\
%-{e}\nabla\cdot\left[n_e\bV_{\!e}\bV_{\!e}+\frac{1}{m_e}\bP\right]
%+{e}\nabla\cdot\left[n_e\bV_{\!i}\bV_{\!i}+\frac{Z}{m_i}\widetilde{\Bbb{P}}_i\right],
%\end{multline*}
%which in turn can be rearranged as
\begin{multline}\label{Burby-ohms}
\left({1}+\frac{Zm_e}{m_i}\right)\bE=
-
\left({1}+\frac{Zm_e}{m_i}\right)\bV_{\!i}\times\bB+\frac{1}{en_e}\left[\bJ\times\bB
-\nabla\cdot\left(\bP
-\frac{Zm_e}{m_i}\widetilde{\Bbb{P}}_i\right)\right]
\\
+\frac{m_e}{e^2n_e}\left[\frac{\partial\bJ}{\partial t}
+\nabla\cdot\left(\bV_{\!i}{\bJ}+{\bJ}\bV_{\!i}-\frac{\bJ\bJ}{en_e}\right)\right]
,
\end{multline}
where Faraday's law can be used to write $\partial\bJ/\partial t=-\mu_0^{-1}\nabla\times\nabla\times\bE$. At this point, upon following a standard procedure in plasma theory, \cite{CJ} neglected all terms of the order of $m_e/m_i$, thereby leading to 
\begin{multline}\label{ohmCJ}
\bE=-\bV_{\!i}\times\bB+\frac{1}{en_e}\left[\bJ\times\bB
-\nabla\cdot\left(\bP
-\frac{Zm_e}{m_i}\widetilde{\Bbb{P}}_i\right)\right]
\\
+\frac{m_e}{e^2n_e}\left[\frac{\partial\bJ}{\partial t}
+\nabla\cdot\left(\bV_{\!i}{\bJ}+{\bJ}\bV_{\!i}-\frac{\bJ\bJ}{en_e}\right)\right],
\end{multline}
where we have recalled that  $\widetilde{\Bbb{P}}_i$ is proportional to the ion mass in order to retain  ion pressure effects. Also,  the relation ${m_e}/{e^2n_e}=\mu_0\delta_e^2$ can be used to rewrite the second line of the equation above in terms of the plasma skin depth.

{ In the present work, we are interested in how electron pressure anisotropy effects manifest in different models. Thus,} we shall study the consequences of using the reduced forms  \eqref{ohmslaw2} and \eqref{ohmCJ} of Ohm's law in the particular case of the Weibel instability \cite{Weibel}. More particularly, we shall consider the implications of both truncated moment models and fully kinetic theories. { Also, special emphasis will be given to the comparison between certain kinetic models and their variational versions, which arise from Hamilton's variational principle \cite{Tronci}.} As we shall see, the approaches based on the simplified Ohm's law \eqref{ohmslaw2} appear unable to capture pressure anisotropy effects without exhibiting physical inconsistencies. While the first part of the paper focuses on moment truncations, the second part is devoted to fully kinetic theories. Finally, the third part shows how quasi-neutral kinetic models based on the generalized Ohm's laws \eqref{Burby-ohms} and \eqref{ohmCJ} appear to recover all the relevant physical features of the Weibel instability.

\section{Moment models\label{sec:moments}}

In order to formulate a simplified model for collisionless reconnection,  \cite{HeWi94c} formulated a hybrid model in which ion kinetics is coupled to a moment truncation of the electron kinetic equation, while the electron momentum equation is replaced by Ohm's law \eqref{ohmslaw2}. The problem of moment truncations is still an active area of research \cite{WaEtAl} dating back to Grad's work \cite{Grad}. In this Section, we linearize the  Hesse-Winske model to study its dispersion relation in the case of the Weibel instability.

\subsection{The Hesse-Winske moment model\label{sec:HW}}

As anticipated above, the Hesse-Winske (HW) model involves a moment truncation of the electron kinetics. More specifically, the electron kinetic equation is truncated to the second-order moment thereby leading to the following equation for the electron pressure (e.g., see equation (2) in \cite{KuHeWi98}):
\begin{multline}\label{pressure-eq}
\frac{\partial\widetilde{\Bbb{P}}_e}{\partial t}+(\bV_{\!e}\cdot\nabla)\widetilde{\Bbb{P}}_e+(\nabla\cdot\bV_{\!e})\widetilde{\Bbb{P}}_e+\widetilde{\Bbb{P}}_e\cdot\nabla\bV_{\!e}+\big(\widetilde{\Bbb{P}}_e\cdot\nabla\bV_{\!e}\big)^T
=\frac{e}{m_e}\left(\bB\times\widetilde{\Bbb{P}}_e-\widetilde{\Bbb{P}}_e\times\bB\right)
.
\end{multline}
This equation neglects heat flux contributions and this approximation may or may not be  physically consistent depending on the case under study. In a series of papers \cite{HeWi94b,HeWi94c,KuHeBi,KuHeWi98,KuHeWi,YiWi,YiWiGaBi,WiHe}, the authors approximated heat flux contributions by an isotropization term involving {\it ad hoc} parameters. However, in this Section we shall continue to discard the heat flux, whose corresponding effects will be completely included in our later discussion of fully kinetic models. We address the reader to Basu's work \cite{Basu} and the more recent { results in} \cite{Ghizzo,SaSaGh} for a complete description of the Weibel instability in terms of kinetic moments. In addition, we point out { that the gyration terms on the right hand side of \eqref{pressure-eq} are discarded in \cite{YiWi,YiWiGaBi}  (strong electron magnetization assumption)}, while these terms are  retained in the present treatment.

The electron pressure dynamics \eqref{pressure-eq} is coupled in the HW model to Faraday's law $\partial\bB/\partial t=-\nabla\times\bE$ and the ion kinetics
\begin{equation}
\label{ion-f}
\frac{\partial f_i}{\partial t}+\bv\cdot\frac{\partial f_i}{\partial \bq}+\frac{Ze}{m_i}\big(\bE+\bv\times\bB\big)\cdot\frac{\partial f_i}{\partial \bv}=0
\,,
\end{equation}
where the electric field is given by Ohm's law in the form \eqref{ohmslaw2}. In addition, quasi-neutrality gives
\begin{equation}\label{neutrality}
Zen_i-en_e=0
\,,\qquad\qquad
Zen_i\bV_{\!i}-en_e\bV_{\!e}=\mu_0^{-1}\nabla\times\bB=\bJ
\,.
\end{equation}
so that $(n_e,\bV_{\!e})$ can be expressed in terms of the ion moments.

Since we are interested in the Weibel instability, we linearize the HW model around a static anisotropic equilibrium of { the type}
\begin{equation}\label{equil}
\bE_0=\bB_0=\bV_{\!e\,0}=\bV_{\!i\,0}=0\,,
\qquad
\Bbb{P}_0=p_\perp\boldsymbol{1}+(p_\parallel-p_\perp)\z\z
\,,\qquad
f_0=f_0(v_\perp^2,v_z^2)
\,,
\end{equation}
and we restrict to consider longitudinal propagation along the wavevector $\bk=k\z$ (here, $\z$ denotes the unit vertical). Notice that we have dropped the species subscripts for convenience and we have retained both electron and ion anisotropies.
 The corresponding dispersion relation is found in Appendix \ref{app:1} and it reads
\begin{equation}\label{dispHW}
\frac{\omega^2}{k^2 v_{e\|}^2}=
1-
\frac{T_\perp^{(e)}}{T_\parallel^{(e)}}
+
k^2\delta^2_e+Z\bmu\left[1+\frac{T_\perp^{(i)}}{T_\|^{(i)}}\,W\!\left(\frac{\omega}{kv_{i\|}}\right)\right]
,
\end{equation}
where $\bmu=m_e/m_i$. 

In order to distinguish the various contributions from the ions and the electrons, it is useful to study the electron Weibel instability and the ion Weibel instability separately. In the first case, one can restrict to an isotropic ion equilibrium, so that $T_\perp^{(i)}=T_\|^{(i)}$. In addition, upon adopting a cold-fluid closure for the ion dynamics one can write $W({\omega}/{kv_{i\|}})\simeq0$ to obtain
\begin{equation}\label{dispHW-e}
\frac{\omega^2}{k^2 v_{\|}^2}=1-\frac{T_\perp^{(e)}}{T_\parallel^{(e)}}
+k^2\delta^2
+
Z\bmu
\,.
\end{equation}
A  detailed discussion of the dispersion relation \eqref{dispHW-e} is presented later in the paper. { For the moment, we remark that $\omega$ is imaginary only in the range $k^2\delta^2<{T_\perp^{(e)}}/{T_\parallel^{(e)}}-Z\bmu-1$, while purely oscillating modes emerge otherwise.}

The ion Weibel instability can be studied in a similar way upon setting $T_\perp^{(e)}=T_\parallel^{(e)}$ in \eqref{dispHW} so that, upon restoring the species index and by denoting by $v_e$ the electron thermal velocity, we have
\begin{equation}\label{dispHW-i}
\frac{\omega^2}{k^2 v_{e}^2}-k^2\delta^2_e
=Z\bmu\left[
1+\frac{T_\perp^{(i)}}{T_\|^{(i)}}\,W\!\left(\frac{\omega}{kv_{i\|}}\right)\right]
.
\end{equation}

Again, this dispersion relation is discussed later in this paper.

\subsection{The effect of Coriolis force terms\label{sec:Cor}}
In \cite{Tronci}, one of us showed how one can neglect the electron mean-flow inertia terms in \eqref{ohmslaw1} by using variational methods based on Hamilton's principle. This approach has the advantage of preserving the total energy and momentum and in recent years there is an increasing amount of work in exploiting this approach for nonlinear plasma modeling \cite{Brizard,CeHoHoMa,HoTr2011,Morrison2005,TrCa}. Essentially, in plasma physics this approach goes back to \cite{Newcomb,Low} and it was later used in \cite{Littlejohn} in his theory of guiding-center motion. When applied to the case under study, this method { produces} Coriolis forces in the electron kinetics that modify the electron pressure dynamics \eqref{pressure-eq} in the HW model as follows:
\begin{multline}\label{pressure-eq2}
\frac{\partial\widetilde{\Bbb{P}}_e}{\partial t}+(\bV_e\cdot\nabla)\widetilde{\Bbb{P}}_e+(\nabla\cdot\bV_e)\widetilde{\Bbb{P}}_e+\widetilde{\Bbb{P}}_e\cdot\nabla\bV_e+\big(\widetilde{\Bbb{P}}_e\cdot\nabla\bV_e\big)^T
\\
+\widetilde{\Bbb{P}}_e\times\left(\frac{e}{m_e}\bB+\boldsymbol\omega_e\right)
-\left(\frac{e}{m_e}\bB+\boldsymbol\omega_e\right)\times\widetilde{\Bbb{P}}_e
=0
\,,
\end{multline}
where $\boldsymbol\omega_e=\nabla\times\bV_{\!e}$ denotes the electron hydrodynamic vorticity. As shown in \cite{Tronci}, the vorticity terms arise by neglecting the electron mean flow inertia after expressing the electron kinetics in the relative frame moving with the Eulerian velocity $\bV_{\!e}$; this takes the dynamics in a non-inertial frame thereby producing Coriolis forces that shift the magnetic field by the electron vorticity. We remark that the terms involving the electron velocity (including the vorticity terms) combine into a fluid transport operator (Lie derivative) so that the electron pressure becomes frozen into the electron mean flow  in the case of strong electron magnetization (so that $ \frac{e}{m_e}\widetilde{\Bbb{P}}_e\times\bB
-\frac{e}{m_e}\bB\times\widetilde{\Bbb{P}}_e\simeq0$). At this point, the Coriolis forces in the electron pressure dynamics lead to a { modified version of the HW model}.

Upon linearizing the modified HW model around the equilibrium \eqref{equil}, one obtains the dispersion relation (see Appendix \ref{app:1})
\begin{equation}\label{dispHW-m}
\frac{\omega^2}{k^2 v_{e\|}^2}=
1-
\frac{T_\perp^{(e)}}{T_\parallel^{(e)}}\left\{
1-k^2\delta^2_e-
\frac{Zm_e}{m_i}\left[1+\frac{T_\perp^{(i)}}{T_\|^{(i)}}\,W\!\left(\frac{\omega}{kv_{i\|}}\right)\right]
\right\}
\,.
\end{equation}

By proceeding analogously to the previous Section, we consider the electron Weibel instability by setting $T_\perp^{(i)}=T_\|^{(i)}$ and $W(\omega/kv_{i\|})\simeq0$ thereby obtaining 
\begin{equation}\label{dispHW-me}
\frac{\omega^2}{k^2 v_{\|}^2}=1-
\frac{T_\perp^{(e)}}{T_\parallel^{(e)}}
+
\frac{T_\perp^{(e)}}{T_\parallel^{(e)}}\left(
k^2\delta^2 +
Z\bmu
\right).
\end{equation}
{ Again, we notice that $\omega$ is imaginary only in the range $k^2\delta^2<1-Z\bmu-{T_\parallel^{(e)}}/{T_\perp^{(e)}}$, while purely oscillating modes emerge otherwise.}

In the same way, we can specialize \eqref{dispHW-m} to the case of the ion Weibel instability. In this case, Coriolis effects become irrelevant and one obtains again equation \eqref{dispHW-i}.

The next section presents a study of equations \eqref{dispHW} and \eqref{dispHW-m} in each considered case.

\subsection{Discussion   on moment models}
Here and in the following discussions, we consider an electron-proton plasma, with typical solar wind parameters.  For comparison, we report the following dispersion relation corresponding to full Maxwell-Vlasov dynamics (see e.g. \cite{GaKa}), as it is obtained by using the exact form of Ohm's law \eqref{Burby-ohms}:
 \begin{equation}\label{equil-kin-MV}
1+\frac{T_\perp^{(e)}}{T_\|^{(e)}}\,W\!\left(\frac\omega{kv_{e\|}}\right)=\frac{\omega^2}{\omega_{pe}^2}-k^2\delta_e^2-Z\bmu \frac{T_\perp^{(i)}}{T_\|^{(i)}}\left[1+W\!\left(\frac\omega{kv_{i\|}}\right)\right].
\end{equation}
Here, the electron Weibel instability is studied by adopting a cold-fluid closure for ion kinetics, so that $T_\perp^{(i)}=T_\|^{(i)}$ and $W(\omega/kv_{i\|})\simeq0$ yield
 \begin{equation}\label{dispMV-e}
1+\frac{T_\perp^{(e)}}{T_\|^{(e)}}\,W\!\left(\frac\omega{kv_{e\|}}\right)=\frac{\omega^2}{\omega_{pe}^2}-k^2\delta_e^2-Z\bmu.
\end{equation}

\begin{figure}[htb!]\center
 \noindent\includegraphics[width=0.8\textwidth]{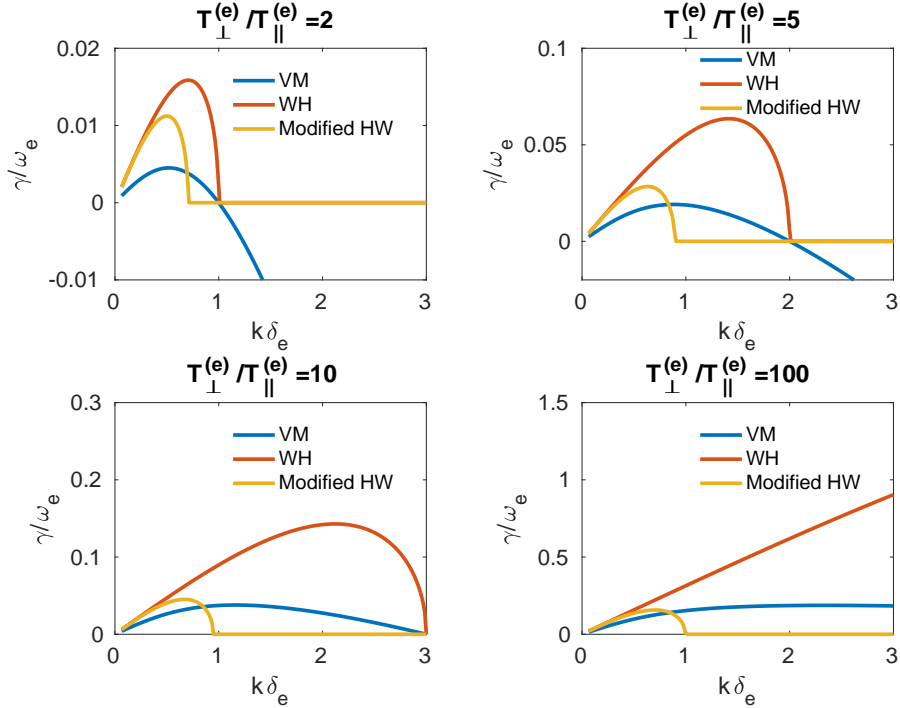}
 \caption{\footnotesize Growth rate for the electron Weibel instability for the HW \eqref{dispHW-e} and modified HW \eqref{dispHW-me} models, for four values of temperature anisotropy $T_\perp^{(e)}/T_\parallel^{(e)}=2, 5, 10, 100$. The blue lines are the reference solutions derived from the Vlasov-Maxwell model, while red and yellow lines are for Eqs.(\ref{dispHW-e}) and (\ref{dispHW-me}), respectively. }\label{fig:HW_electron}
\end{figure}

\begin{figure}[htb!]\center
 \noindent\includegraphics[width=0.8\textwidth]{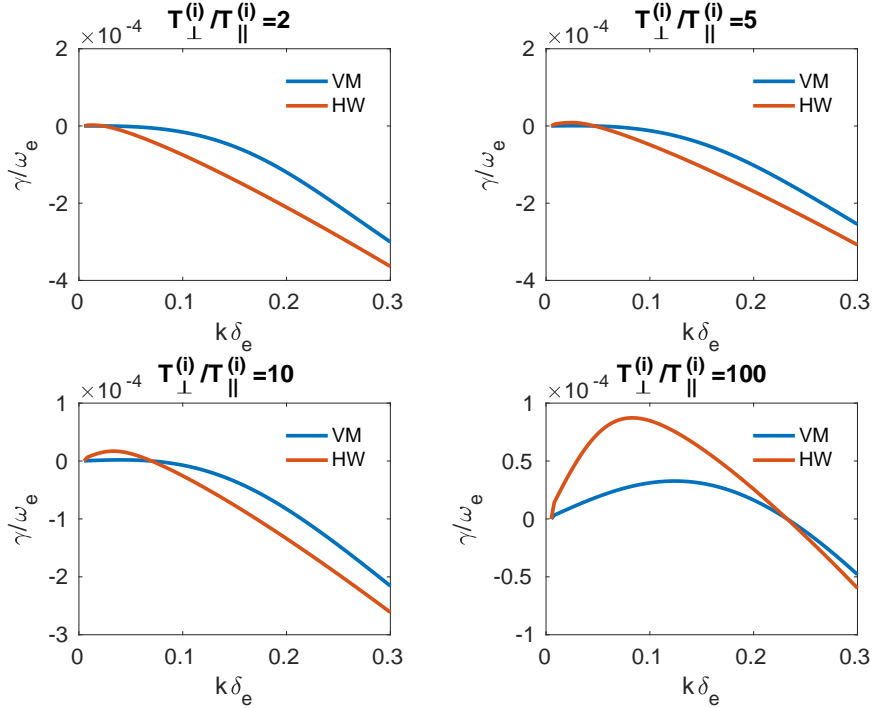}
 \caption{\footnotesize Growth rate for the ion Weibel instability for the HW \eqref{dispHW} model, for four values of temperature anisotropy $T_\perp^{(i)}/T_\parallel^{(i)}=2, 5, 10, 100$. The blue lines are the reference solutions derived from the Vlasov-Maxwell model, while red lines are for Eq.(\ref{dispHW}). In this case the modified HW model (\ref{dispHW-m}) yields identical results.}\label{fig:HW_ion}
\end{figure}

Figures \ref{fig:HW_electron} and \ref{fig:HW_ion} show the dispersion relations for electron and ion Weibel instabilities, respectively. The ratio between electron thermal velocity { (in the cold direction) }and speed of light $v_e/c=0.0318$ (this is also the ratio between Debye length and electron inertial length). The mass ratio is physical $m_i/m_e=1836$. 
The four panels are for values of temperature anisotropy equal to 2, 5, 10, and 100. 
The blue lines show the reference solutions derived from the Vlasov-Maxwell model { \eqref{dispMV-e} (involving a cold-fluid closure for ion kinetics)}, while red lines are for Eq.(\ref{dispHW-e}).  In Figure \ref{fig:HW_electron} one can notice how the HW model yields much larger growth rates than the correct values. The results for the modified HW model \eqref{dispHW-me} are shown in yellow. They partially correct the discrepancies with the full Vlasov-Maxwell model, but they are still unsatisfactory, especially for wavevectors larger than the inverse electron inertial length. 

As we mentioned, the Coriolis effects are irrelevant for the case of the ion Weibel instability.   
In this case, the reference Vlasov solution is obtained in  Figure \ref{fig:HW_ion} by solving the dispersion relation
\begin{equation}\label{equil-kin-MV3}
1+\frac{k^2v_{e\parallel}^2}{2\omega^2}=\frac{\omega^2}{\omega_{pe}^2}-k^2\delta_e^2-Z\bmu \frac{T_\perp^{(i)}}{T_\|^{(i)}}\left[1+W\!\left(\frac\omega{kv_{i\|}}\right)\right].
\end{equation}
This is derived upon adopting a warm-fluid closure for electron kinetics, that is by inserting $T_\perp^{(e)}=T_\|^{(e)}$ and $W({\omega}/{kv_{e\|}})\simeq(1/2){k^2v_{e\parallel}^2}/{\omega^2}$ in \eqref{equil-kin-MV}. Interestingly, for the ion Weibel instability the discrepancies between the HW and VM models are already significant for $k\delta_e<0.1$.

\section{Electron inertia in fully kinetic theories}

{ While the results in the previous sections were obtained by using moment truncations, one is led to ask about the effects arising from higher moments. In order to address this point, this Section presents two different ways to neglect the electron mean flow inertia in a fully kinetic theory, in such a way that all higher moments are fully considered. We remark that this is an unprecedented approach in the plasma physics literature, with the only exception of \cite{Tronci}.
By following the discussion therein}, we remark that it may not be convenient to implement  this approximation directly in the electron kinetic equation
\begin{equation}
\label{electron-f}
\frac{\partial f_e}{\partial t}+\bv\cdot\frac{\partial f_e}{\partial \bq}-\frac{e}{m_e}\big(\bE+\bv\times\bB\big)\cdot\frac{\partial f_e}{\partial \bv}=0
\,.
\end{equation}
Indeed, doing this would generate questions of compatibility between the above electron kinetics and  the reduced form of Ohm's law \eqref{ohmslaw2}, which we want to adopt throughout this Section as a first step in neglecting the electron mean flow inertia. Before making any assumption, it is instead convenient to express electron kinetics in the mean-flow frame by introducing the coordinate $\bc=\bv-\bV_{\!e}$ and looking at the dynamics for the relative distribution
\[
\mathsf{f}_e(\bx,\bc,t)=f_e(\bx,\bc+\bV_{\!e},t)
\,,
\]
that is
\begin{equation}\label{el-rel}
\frac{\partial \mathsf{f}_e}{\partial t}+(\bc+\bV_{\!e})\cdot\frac{\partial \mathsf{f}_e}{\partial \bq}-\left\{\frac{D\bV_{\!e}}{D t}+(\bc\cdot\nabla)\bV_{\!e}+\frac{e}{m_e}\Big[\bE+(\bc+\bV_{\!e})\times\bB\Big]\right\}\cdot\frac{\partial \mathsf{f}_e}{\partial \bc}=0
\,.
\end{equation}
In turn, this kinetic equation is accompanied by Amp\`ere's law and Faraday's law. At this stage, one still needs a closure for the electric field, which can be obtained by writing Ohm's law. The latter arises from taking the first moment of \eqref{el-rel} and by using the constraint $\int\!\bc\,\mathsf{f}_e(\bc)\,\de\bc=0$; this process leads to equation \eqref{ohmslaw}. So far, no approximation was performed and the mean-flow electron inertia is still fully retained, as it is made explicit by multiplying \eqref{el-rel} by $m_en_e$. Indeed, we notice that the first term in the acceleration field multiplying ${\partial \mathsf{f}_e}/{\partial \bc}$ in \eqref{el-rel} is precisely the term neglected in Ohm's law \eqref{ohmslaw1} to obtain its reduced form \eqref{ohmslaw2}. This acceleration term can also be expanded as 
\begin{equation}\label{V-exp}
\frac{D\bV_{\!e}}{D t}=\frac{\partial\bV_{\!e}}{\partial t}-\bV_{\!e}\times(\nabla\times\bV_{\!e})+\frac12\nabla|\bV_{\!e}|^2,
\end{equation}
which evidently corresponds to a superposition of inertial forces excerpted by the mean flow on the particles moving in the relative frame.

In the next Section, we shall present two different possible strategies for implementing the assumption of negligible electron mean-flow inertia. While the first approach is direct and involves the equations of motion, the second approach is based on variational methods and it involves Hamilton's principle. Although the second approach removes some of the inconsistencies emerging from the first, both methods appear to be unsatisfactory for a complete description of the Weibel instability.

\subsection{Removing the electron inertia\label{sec:kin-HW}}

A first approach to neglect electron inertia consists of simply removing the  term $-{D\bV_{\!e}}/{D t}$ in \eqref{el-rel}, thereby leading to the modified electron equation
\begin{equation}\label{el-rel-HW}
\frac{\partial \mathsf{f}_e}{\partial t}+(\bc+\bV_{\!e})\cdot\frac{\partial \mathsf{f}_e}{\partial \bq}-\left\{\bc\cdot\nabla\bV_{\!e}+\frac{e}{m_e}\Big[\bE+(\bc+\bV_{\!e})\times\bB\Big]\right\}\cdot\frac{\partial \mathsf{f}_e}{\partial \bc}=0
\,.
\end{equation}
Although this equation retains the acceleration term $-\bc\cdot\nabla\bV_{\!e}$, inertial forces are only partially considered since the term \eqref{V-exp} has been entirely neglected. At this point, one can easily take the first moment of \eqref{el-rel-HW}, so that using the constraint $\int\!\bc\,\mathsf{f}_e(\bc)\,\de\bc=0$ leads to the reduced Ohm's law \eqref{ohmslaw2} and a fully kinetic model is formulated by  using ion kinetics \eqref{ion-f}, along with Amp\`ere's and Faraday's laws.

The model obtained in this way provides the basis for the HW moment model in Section \ref{sec:HW}, except that the HW model invokes the quasi-neutrality conditions \eqref{neutrality}. 
The idea of using quasi-neutrality in a fully kinetic model is not new. In later sections, we shall show how the quasi-neutrality assumption can be used successfully in fully kinetic theories, although it requires extra care. However, for the purpose of this Section, we shall keep assuming quasi-neutrality in the present discussion. Thus, Amp\`ere's law in \eqref{neutrality} can be used to eliminate entirely the variable $\bV_{\!e}$ in favour of the ion velocity $\bV_{\!i}$, as it is computed from \eqref{ion-f}.

Combining \eqref{ion-f}, \eqref{el-rel-HW}, \eqref{neutrality}, and Faraday's law yields a fully kinetic model, whose moment truncation to second-order yields exactly the HW moment model from Section \ref{sec:HW}. For later reference, we { shall refer to this as} the {\it HW kinetic model}. Then, one would hope that completing the HW moment model by retaining fully kinetic effects (while still neglecting electron mean-flow inertia) could capture more { physics}. As we shall see, this may not always be true and we explain this below by considering again the case of the Weibel instability. 

Here, we linearize the HW kinetic model around the bi-Maxwellian equilibrium
\begin{equation}\label{equil-kin}
\bE_0=\bB_0=0
\,,\qquad\qquad
f_0=f_0(v_\perp^2,v_z^2)
\,,\qquad\qquad
\mathsf{f}_0=\mathsf{f}_0(c_\perp^2,c_z^2)
\,,
\end{equation}
where $f_0$ and $\mathsf{f}_0$ denote the ionic and electronic equilibrium, respectively. As shown in Appendix \ref{app:2}, we obtain the dispersion relation 
\begin{equation}\label{dispHWkin}
1+\frac{T_{\perp}^{(e)}}{T_{\|}^{(e)}}\,W\!\left(\frac{\omega}{kv_{e\parallel}}\right)
=\left\{k^2\delta^2_e+{Z\bmu}\left[1+\frac{T_\perp^{(i)}}{T_\|^{(i)}}\,W\!\left(\frac{\omega}{kv_{i\|}}\right)\right]\right\}W\!\left(\frac{\omega}{kv_{e\parallel}}\right).
\end{equation}
In order to study the electron Weibel instability, we follow the approach in Section \ref{sec:HW} and adopt a cold-fluid closure for the ions by setting $T_\perp^{(i)}=T_\|^{(i)}$ and $W({\omega}/{kv_{i\|}})\simeq0$. This yields
\begin{equation}\label{dispHWkin-e}
1+\frac{T_{\perp}^{(e)}}{T_{\|}^{(e)}}\,W\!\left(\frac{\omega}{kv_{e\parallel}}\right)
=(k^2\delta^2_e+{Z\bmu})\,W\!\left(\frac{\omega}{kv_{e\parallel}}\right).
\end{equation}
On the other hand, the ion Weibel instability requires special care since Ohm's law \eqref{ohmslaw2} requires pressure to balance the Lorentz force in electron dynamics. { Indeed, as one can see especially in equation \eqref{lin-ohm} in Appendix \ref{app:1}, adopting a cold-fluid closure for electron dynamics would lead to consistency issues}. However, a warm fluid closure can be performed by setting $T_\perp^{(e)}=T_\|^{(e)}$ and $W({\omega}/{kv_{e\|}})\simeq(1/2){k^2v_{e\parallel}^2}/{\omega^2}$ { so that \eqref{dispHWkin} becomes}
\begin{equation}\label{dispHWkin-i}
\frac{2\omega^2}{k^2v_{e\parallel}^2}-k^2\delta^2_e
={Z\bmu}\left[1+\frac{T_\perp^{(i)}}{T_\|^{(i)}}\,W\!\left(\frac{\omega}{kv_{i\|}}\right)\right]-1
\,.
\end{equation}
The contribution of the heat flux and higher moments can be understood by comparing the above equation to the corresponding equation \eqref{dispHW-i} for the HW moment model.

While the discussion of the dispersion relations \eqref{dispHWkin-e} and \eqref{dispHWkin-i} is left for later discussion, the next Section aims at extending the modified HW moment model from Section \ref{sec:Cor} to a fully kinetic theory.

\subsection{Coriolis force effects\label{sec:kin-Tronci}}
A modified version of the HW kinetic model was presented in \cite{Tronci} (see equations (1)-(5) therein), by exploiting variational techniques based on Hamilton's principle. As discussed in Section \ref{sec:Cor}, this approach produces the Coriolis force terms appearing in equation \eqref{pressure-eq2}. In the fully kinetic treatment, the same approach leaves  \eqref{ion-f}, \eqref{neutrality}, and Faraday's law unchanged while  \eqref{el-rel-HW} is modified as follows:
\begin{equation}\label{el-rel-Tronci}
\frac{\partial \mathsf{f}_e}{\partial t}+(\bc+\bV_{\!e})\cdot\frac{\partial \mathsf{f}_e}{\partial \bq}-\left\{\bc\cdot\nabla\bV_{\!e}+\bc\times\nabla\times\bV_{\!e}+\frac{e}{m_e}\Big[\bE+(\bc+\bV_{\!e})\times\bB\Big]\right\}\cdot\frac{\partial \mathsf{f}_e}{\partial \bc}=0
\,.
\end{equation}
Evidently, this differs from \eqref{el-rel-HW} by the Coriolis acceleration term $\bc\times\nabla\times\bV_{\!e}$. As discussed in \cite{Tronci}, this term appears from the variational approach due to the fact that the change of frame performed to express the electron kinetics in the mean-flow frame affects the Lorentz force term, which now is written in terms of the effective magnetic field $\bB+m_e\boldsymbol\omega_e/e$. This is a typical feature of electrodynamics in non-inertial frames, as explained in \cite{ThMc}. Notice that the Coriolis acceleration term is absent in \eqref{el-rel}, which also means that this term is produced to guarantee a consistent  force balance after the mean-flow inertia term $D\bV_{\!e}/Dt$ is dropped in \eqref{el-rel}. At this point, the modified HW { kinetic} model is given by \eqref{el-rel-Tronci},  \eqref{ion-f}, \eqref{neutrality}, and Faraday's law.

For comparison with the HW kinetic model in the previous Section, we study the effect of Coriolis forces by  considering again the Weibel instability. Then, we linearize the modified HW kinetic system around the equilibrium \eqref{equil-kin} to obtain the dispersion relation (see Appendix \ref{app:2})
\begin{equation}\label{dispHWkin-m}
1+\frac{T_{\perp}^{(e)}}{T_{\|}^{(e)}}W\!\left(\frac{\omega}{kv_{e\parallel}}\right)
=\frac{T_{\perp}^{(e)}}{T_{\|}^{(e)}}\left\{k^2\delta^2_e+{Z\bmu}\left[1+\frac{T_\perp^{(i)}}{T_\|^{(i)}}\,W\!\left(\frac{\omega}{kv_{i\|}}\right)\right]\right\}W\!\left(\frac{\omega}{kv_{e\parallel}}\right).
\end{equation}
By following the approach in the previous Sections, we restrict to consider the electron Weibel instability by adopting a cold-fluid closure for the ions. This yields
\begin{equation}\label{dispHWkin-m-e}
\frac{T_{\|}^{(e)}}{T_{\perp}^{(e)}}+W\!\left(\frac{\omega}{kv_{e\parallel}}\right)
=(k^2\delta^2_e+{Z\bmu})\,W\!\left(\frac{\omega}{kv_{e\parallel}}\right).
\end{equation}
On the other hand, upon assuming a warm-fluid closure for the electrons by replacing $T_\perp^{(e)}=T_\|^{(e)}$ and $W({\omega}/{kv_{e\|}})\simeq(1/2){k^2v_{e\parallel}^2}/{\omega^2}$ in \eqref{dispHWkin-m}, we obtain the same dispersion relation \eqref{dispHWkin-i} for the ion Weibel instability.

\subsection{Discussion on   kinetic models with inertialess electrons}

In first instance, this Section compares the dispersion relations \eqref{dispHWkin-e} and \eqref{dispHWkin-m} for the electron Weibel instability with the corresponding result \eqref{dispMV-e}
%\begin{equation}\label{dispMV-e}
%1+\frac{T_{\perp}^{(e)}}{T_{\|}^{(e)}}\, W\!\left(\frac{\omega}{kv_{e\parallel}}\right)
%=
%\frac{\omega^2}{\omega_{pe}^2}-{k^2\delta^2} -Z\bmu 
%\end{equation}
{ for the case of the Maxwell-Vlasov system for cold-fluid ions}. A typical limit that is often used to study the Weibel instability is
\[
\omega\ll kv_{e\parallel}
\]
so that $W({\omega}/{kv_{e\parallel}})\simeq-1-i\sqrt\pi\,{\omega}/{kv_{e\parallel}}$. In this limit, equations \eqref{dispHWkin-e} and \eqref{dispHWkin-m} become (upon dropping the species superscript for convenience)
\begin{equation}\label{mario1}
-i{\omega}=\frac{kv_{\parallel}}{\sqrt\pi}\left(1+\frac{1}{T_{\perp}/T_{\|}-k^2\delta^2-Z\bmu}\right)
\end{equation}
and 
\begin{equation}\label{mario2}
-i{\omega}=\frac{kv_{\parallel}}{\sqrt\pi}\left(1-\frac{T_{\|}}{T_{\perp}} \frac{1}{1-k^2\delta^2-Z\bmu} \right),
\end{equation}
respectively. On the other hand, upon assuming $\omega\ll\omega_{p}$, equation \eqref{dispMV-e} becomes
\begin{equation}\label{mario3}
-i{\omega}=\frac{kv_{\parallel}}{\sqrt\pi}\left[1 -\frac{T_{\|}}{T_{\perp}}(k^2\delta^2+Z\bmu+1)\right]
\end{equation}
Now, we observe that in the limit $k\delta\ll1$ the results in \eqref{mario2} and \eqref{mario3} coincide thereby showing that the variational model from Section \ref{sec:Cor} agrees well with Maxwell-Vlasov dynamics for lenghtscales much bigger than the skin depth. In turn, in the same limit $k\delta\ll1$ \eqref{mario1}  disagrees with the Maxwell-Vlasov result \eqref{mario3}  with growing anisotropies.

\begin{figure}[htb!]\center
 \noindent\includegraphics[width=0.6\textwidth]{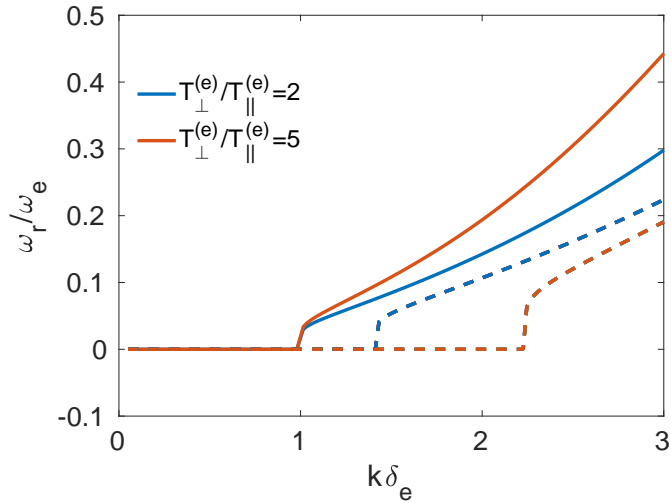}
 \caption{\footnotesize Real part of the frequency for the electron Weibel instability for the modified HW kinetic model \eqref{dispHWkin-m-e} (solid lines) and for the KW kinetic model \eqref{dispHWkin-e} (dashed lines), for values of temperature anisotropy $T_\perp^{(e)}/T_\parallel^{(e)}=2, 5$.}\label{fig:HW_kinetic_realpart}
\end{figure}

However, both results \eqref{mario1} and \eqref{mario2} suffer from the important drawback that a vertical asymptote emerges in the growth rate as lengthscales approach the skin depth. We remark that the assumption $\omega\ll kv_{e\parallel}$ is no longer valid near and after the asymptote and so the dispersion relation needs to be solved numerically, as presented below.  After the asymptotes, for both kinetic models the least damped mode is { not the Weibel mode, but} one with non-zero real frequency, hence yielding a completely different result from the  Maxwell-Vlasov theory, in which the Weibel (purely damping) mode is dominant.
\begin{figure}[htb!]\center
 \noindent\includegraphics[width=0.8\textwidth]{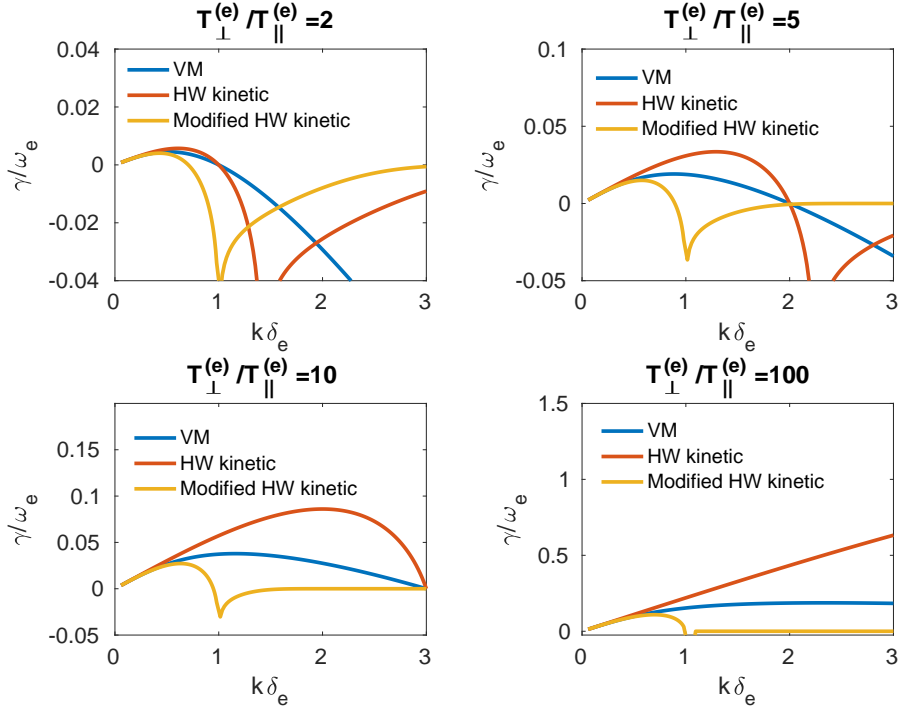}
 \caption{\footnotesize Growth rate for the electron Weibel instability for the HW kinetic \eqref{dispHWkin-e} and modified HW kinetic \eqref{dispHWkin-m-e} models, for four values of temperature anisotropy $T_\perp^{(e)}/T_\parallel^{(e)}=2, 5, 10, 100$. The blue lines are the reference solutions derived from the Vlasov-Maxwell model, while red and yellow lines are for Eqs.(\ref{dispHWkin-e}) and (\ref{dispHWkin-m-e}), respectively. }\label{fig:HW_kinetic_electron}
\end{figure}
Figure \ref{fig:HW_kinetic_electron} shows the dispersion relations for the electron instability, for four values of temperature anisotropy $T_\perp^{(e)}/T_\parallel^{(e)}=2, 5, 10, 100$.
The blue lines are the reference solutions derived from the Vlasov-Maxwell model \eqref{dispMV-e}, while red and yellow lines are for the HW kinetic (\ref{dispHWkin-e}) and modified HW kinetic (\ref{dispHWkin-m-e}) models, respectively. 
The aforementioned asymptote for the reduced models is clearly visible, with the distinguishing features that while it always occurs at $k\delta=1$ for the modified HW model, it becomes a function of anisotropy for the HW model.
Both models present large discrepancies with respect to the full Vlasov-Maxwell solution, with the wave-vector approaching the inverse electron inertial length.
Figure \ref{fig:HW_kinetic_realpart} shows the real frequency of the least damped mode for the modified HW kinetic model (solid lines) and for the HW kinetic model (dashed lines). 
\begin{figure}[h!]\center
 \noindent\includegraphics[width=0.8\textwidth]{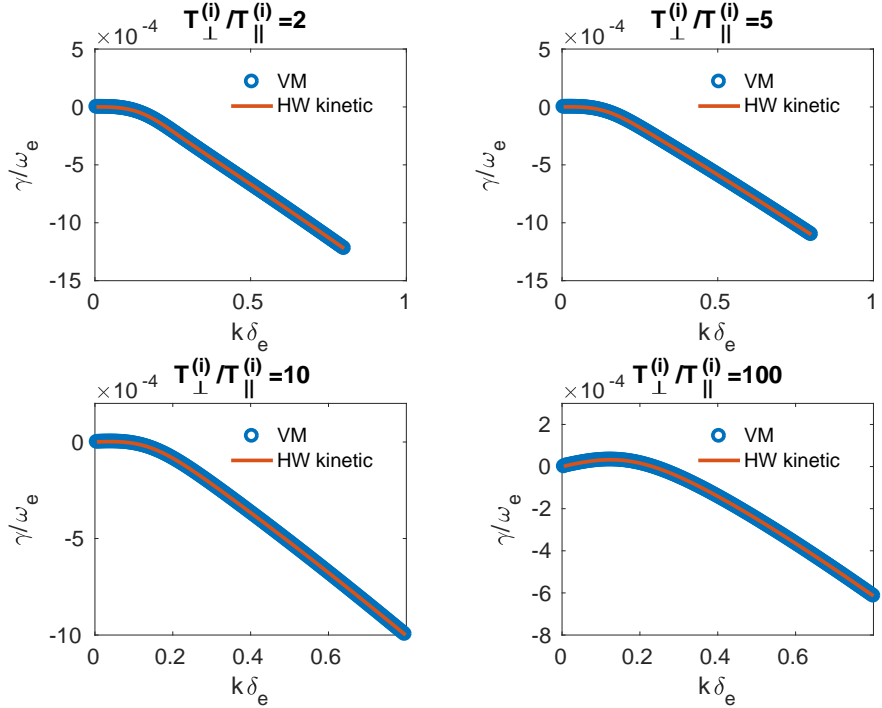}
 \caption{\footnotesize Growth rate for the ion Weibel instability for the HW kinetic model \eqref{dispHWkin}, for four values of temperature anisotropy $T_\perp^{(i)}/T_\parallel^{(i)}=2, 5, 10, 100$. The blue circles denote the reference solutions derived from the Vlasov-Maxwell model, while red lines are for Eq.(\ref{dispHWkin}).}\label{fig:HW_kinetic_ion}
\end{figure}

Once again, in contrast with the correct VM solution, the mode's real frequency becomes non-zero after the respective asymptote.  
The ion Weibel instability is, on the contrary, well captured by both models. This is shown in Figure \ref{fig:HW_kinetic_ion}. 
Once again, in this case the Coriolis correction does not play any role and the two models become identical. One can notice that, for any value of temperature anisotropy, the solutions are indistinguishable from the correct Vlasov-Maxwell results, which are obtained from the dispersion relation \eqref{equil-kin-MV3} (adopting a warm-fluid closure for the electrons). In some sense, this is not surprising, since ion kinetics is not subject to any approximation in either the HW kinetic model and its modified variant.

\section{Quasi-neutral Vlasov theories}

We have shown that all the moment models and fully kinetic theories considered so far and aiming at neglecting the electron inertia in Ohm's law \eqref{ohmslaw1} suffer from different drawbacks. { More specifically, in the nonlinear regime unphysical modes with $k\delta>1$ may be excited even if  $k\delta\ll1$ at the initial time.} Even the variational approach in \cite{Tronci}, while correcting certain discrepancies in the electron Weibel instability and retaining the full physics of the ion Weibel instability, would need an appropriate numerical filtering to prevent the dynamics from introducing lengthscales of the order of { (or smaller than)} the electron skin depth. 
On the other hand, the analysis performed so far also posed an alternative question about the validity of the quasi-neutral limit in fully kinetic theories. Indeed, the assumption of quasi-neutrality \eqref{neutrality} was used throughout all the discussion thereby leading to the question whether { quasi-neutrality} may also produce consistency issues when implemented in a fully kinetic theory. A first answer to this question was provided by Cheng and Johnson in \cite{CJ}, where quasi-neutrality was assumed in the Maxwell-Vlasov system, along with the generalized Ohm's law \eqref{ohmCJ}. In this approach, all terms of the order $m_e/m_i$ are considered irrelevant and thus are ignored. On the other hand, these terms were considered in more recent work by the authors \cite{TrCa}, where quasi-neutrality was invoked at the level of Hamilton's variational principle. The model in \cite{TrCa} was dubbed the \emph{neutral Vlasov model}.

In the following Sections, we present both the Cheng-Johnson (CJ) model \cite{CJ} and the neutral Vlasov model \cite{TrCa}. As we shall see, both models reproduce faithfully the physics of both ion and electron Weibel instabilities. In addition, we shall see how Amp\`ere's current balance may play a crucial role in preserving quasi-neutrality at all times; this point is of particular interest for the CJ model, where the exact current balance is lost, while it is retained by the \emph{Neutral Vlasov model}.

\subsection{The Cheng-Johnson model\label{sec:CJ}}
As anticipated above, \cite{CJ} were the first to design an alternative fully kinetic model in the quasi-neutral limit. More specifically, they expanded  Ohm's law \eqref{ohmslaw} by using Ampe\`ere's and Faraday's laws to obtain  \eqref{Burby-ohms}. Then, after assuming quasi-neutrality to write $\bJ=\mu_0^{-1}\nabla\times\bB$, they neglected all terms of the order of $m_e/m_i$. This process leads to the reduced form of Ohm's law in \eqref{ohmCJ}, which is then accompanied by the kinetic equations \eqref{ion-f} and \eqref{electron-f}, the quasi-neutrality conditions \eqref{neutrality}, and Faraday's law. We remark that originally the Cheng-Johnson (CJ) model was called {\it kinetic-multifluid system} because the kinetic equation for each species was written to accompany the equation for its first moment. However, this is totally equivalent to retaining only the kinetic equations.

In order to compare the CJ model to the systems formulated in the previous Section, we studied the Weibel instability by linearizing again around the equilibrium \eqref{equil-kin}.  Upon linearizing the CJ model around the bi-Maxwellian equilibrium
\begin{equation}\label{equil-kin-CJ}
\bE_0=\bB_0=0
\,,\qquad\qquad
f_{s0}=f_{s0}(v_\perp^2,v_z^2)
\end{equation} 
(where the subscript $s$ refers to the particle species), we obtain the dispersion relation (see Appendix \ref{app:3})
\begin{equation}\label{disprel-CJ}
1+\frac{T_\perp^{(e)}}{T_\|^{(e)}}W\!\left(\frac\omega{kv_{e\|}}\right)=-k^2\delta_e^2-Z\bmu \frac{T_\perp^{(i)}}{T_\|^{(i)}}W\!\left(\frac\omega{kv_{i\|}}\right).
\end{equation}
{ The above dispersion relation can be compared directly with the relation \eqref{equil-kin-MV} that is obtained for full Maxwell-Vlasov dynamics.}
As expected, both electron and ion Weibel instabilities are reproduced by the CJ model in exceptional agreement with the full Maxwell-Vlasov theory.

A point about the CJ model that was omitted so far { (while it deserves some attention)} is that the quasi-neutrality conditions \eqref{neutrality} are not preserved in time exactly, as it can be verified by a direct calculation. More specifically, one can ask if the electron velocity as computed from the first moment of \eqref{electron-f} is compatible with the corresponding expression arising from \eqref{neutrality}. In order to provide the answer to this question, we  use \eqref{neutrality} and the first moment equation of \eqref{ion-f} to rewrite \eqref{ohmCJ} as
\begin{equation}\label{ohmCG-mod}
 \bE=-\bV_{\!i}\times\bB+\left(1+\frac{Z\bmu}{1-Z\bmu}\right)\left(\frac1{e n_e}\,\bJ\times\bB-\frac1{e n_e}\nabla\cdot\bP-\frac{m_e}e\frac{D\bV_{\!e}}{Dt}\right).
\end{equation}
Then, we use \eqref{ohmCG-mod} as the CJ closure equation for the electric field in \eqref{electron-f}. Taking the first moment (here denoted by $\bK_e$) of the resulting electron kinetic equation yields
\begin{multline*}
\frac{\partial}{\partial t}\left(\bK_e-n_e\bV_{\!e}\right)+\bV_e\cdot\nabla\left(\bK_e-n_e\bV_{\!e}\right)
+\nabla\cdot\left(\bV_{\!e\,}\bK_e-n_e\bV_{\!e}\bV_{\!e}\right)
\\
=\left(\bK_e-n_e\bV_{\!e}\right)\times\left(\boldsymbol\omega_e-\frac{e}{m_e}\bB\right)
+\frac{Z\bmu}{1-Z\bmu}\left(\frac1{m_e}\,\bJ\times\bB-\frac1{m_e}\nabla\cdot\bP-n_e\frac{D\bV_{\!e}}{Dt}\right),
\end{multline*}
where we recall that $\boldsymbol\omega_e=\nabla\times\bV_{\!e}$ is the electron vorticity and $(n_e,\bV_{\!e})$ are expressed by using \eqref{neutrality}. Then, we conclude that the consistency relation $\bK_e=n_e\bV_{\!e}$ fails to be preserved in time.
This is a fundamental consistency issue that is intrinsic to the model and can lead to different drawbacks beyond the linear analysis of the Weibel instability. For example, charge conservation is dramatically affected, thereby preventing neutrality from being satisfied at all times. In the next Section, we show how this issue is solved by simply retaining all terms in the exact Ohm's law   \eqref{Burby-ohms}.

\subsection{The neutral Vlasov model}
Recently, upon retaining all terms in the exact Ohm's law \eqref{Burby-ohms}, we  showed \cite{TrCa} how the quasi-neutral limit can be consistently implemented in the Maxwell-Vlasov system both directly (by formally letting $\epsilon_0\to0$) and in Hamilton's variational principle: a comparison with the full Maxwell-Vlasov system was presented and good agreement was found in the linear case for both Alfv\`en and Whistler modes at different angles of propagation.  Later, \cite{Burby} showed how this model also possesses a Hamiltonian structure, while \cite{DeDeDo} provided an alternative mathematical footing by exploiting scaling and asymptotic techniques in the case of one particle species (by preventing ion motion). In the same work, the question of the numerical implementation was also discussed.
As it was presented in \cite{TrCa}, the quasi-neutral model  consists of the kinetic equations \eqref{ion-f} and \eqref{electron-f}, the quasi-neutrality conditions \eqref{neutrality}, Faraday's law, and Ohm's law \eqref{ohmslaw}. Here, the electron velocity is expressed in terms of the ion velocity by using Amp\`ere's law. 

A question that emerged at the end of Section \ref{sec:CJ} concerned the possibility of consistency issues precisely with the second { equation} in \eqref{neutrality}. Again, one asks if the electron velocity as computed from the first moment of \eqref{electron-f} is compatible with the corresponding expression arising from \eqref{neutrality}. A positive answer can be found by following a similar procedure as in the previous Section. First, one replaces \eqref{ohmslaw1}  in \eqref{electron-f} and then one takes the first moment  of the resulting kinetic equation. As a result, one obtains
\begin{multline*}
\frac{\partial}{\partial t}\left(\bK_e-n_e\bV_{\!e}\right)+\bV_e\cdot\nabla\left(\bK_e-n_e\bV_{\!e}\right)
+\nabla\cdot\left(\bV_{\!e\,}\bK_e-n_e\bV_{\!e}\bV_{\!e}\right)
\\
=\left(\bK_e-n_e\bV_{\!e}\right)\times\left(\boldsymbol\omega_e-\frac{e}{m_e}\bB\right),
\end{multline*}
where we recall that $\boldsymbol\omega_e=\nabla\times\bV_{\!e}$ is the electron vorticity and $(n_e,\bV_{\!e})$ are expressed by using \eqref{neutrality}. { Then, we conclude that if $\bK_e=n_e\bV_{\!e}$ is verified initially, then it stays so at all times.
%: one one hand, this is not surprising as the mathematical consistency of the model is ensured by the variational structure, while on the other hand it is perhaps explanatory to have the explicit dynamics of the momentum difference $\bK_e-n_e\bV_{\!e}$.

As a further remark on the neutral Vlasov model, we notice} that Ohm's law \eqref{ohmslaw} is not suitable for the numerical implementation, since the electron mean-flow inertia produces an explicit time derivative in the expression of the electric field. In his thesis, Burby expanded \eqref{ohmslaw} by using \eqref{neutrality} to obtain \eqref{Burby-ohms} in the form
\begin{multline}
\left({1}+\frac{Zm_e}{m_i}\right)\bE+\frac{m_e}{\mu_0e^2n_e}\nabla\times\nabla\times\bE=
-
\left({1}+\frac{Zm_e}{m_i}\right)\bV_{\!i}\times\bB
\\
+\frac{1}{en_e}\left[\bJ\times\bB
-\nabla\cdot\left(\bP
-\frac{Zm_e}{m_i}\widetilde{\Bbb{P}}_i\right)\right]
+\frac{m_e}{e^2n_e}\left[\nabla\cdot\left(\bV_{\!i}{\bJ}+{\bJ}\bV_{\!i}-\frac{\bJ\bJ}{en_e}\right)\right]
,
\end{multline}
 with $\bJ=\mu_0^{-1}\nabla\times\bB$. While on one hand this eliminates the time derivative in the closure for the electric field, one the other hand it involves inverting the operator $\psi+\curl^2$ for some function $\psi(\bx)$. (Here we shall not dwell upon the question of the numerical costs involved in inverting this operator). We remark that here we did not set $\nabla\cdot\bE$ to zero, as there is absolutely no reason for this { to hold: indeed,} quasi-neutrality is obtained by letting $\epsilon_0\to0$ in Gauss' law, while no hypothesis is made on $\nabla\cdot\bE$.
 
 The linear stability of the Neutral Vlasov model can be easily studied since, as already noticed in \cite{TrCa} it suffices to take the limit $\epsilon_0\to0$ in the standard dispersion relation for the Maxwell-Vlasov system. 
Then, for example, the case of the Weibel instability can be studied by simply discarding the term $\omega^2/\omega_{pe}^2$ in \eqref{equil-kin-MV} so that
\begin{equation}\label{disprel-NV}
1+\frac{T_\perp^{(e)}}{T_\|^{(e)}}\,W\!\left(\frac\omega{kv_{e\|}}\right)=-k^2\delta_e^2-Z\bmu \frac{T_\perp^{(i)}}{T_\|^{(i)}}\left[1+W\!\left(\frac\omega{kv_{i\|}}\right)\right].
\end{equation}
In the next Section we show the results obtained with the Neutral Vlasov model for both the electron and the ion Weibel instabilities.

\subsection{Discussion   on quasi-neutral kinetic models}
This Section compares the dispersion relations for the ion and electron Weibel instability derived from the CJ and the neutral Vlasov model. 
\begin{figure}[htb!]
\begin{center}
 \noindent\includegraphics[width=0.8\textwidth]{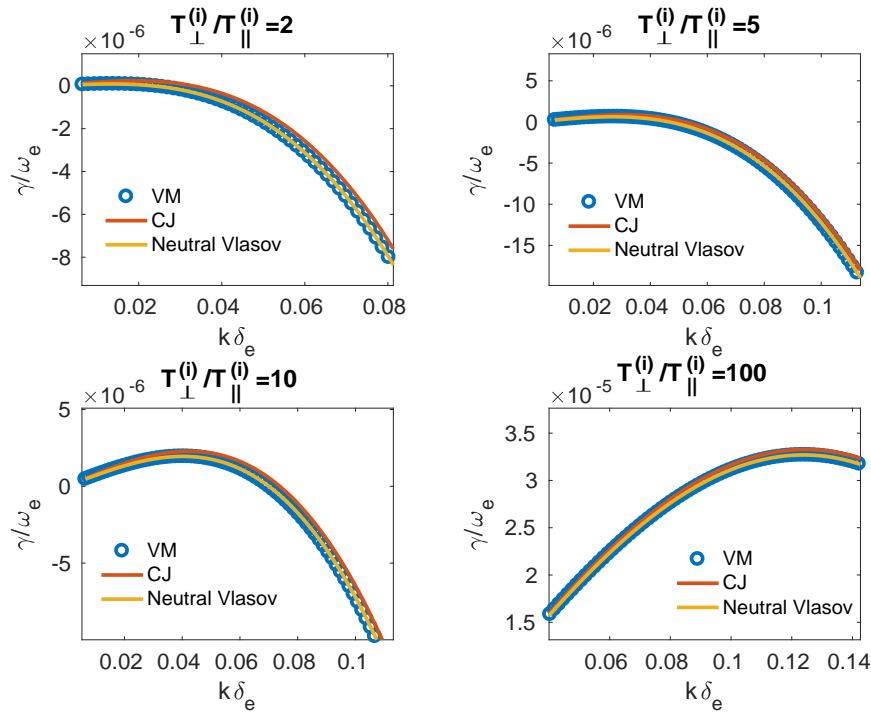}
 \caption{\footnotesize Growth rate for the ion Weibel instability for four values of temperature anisotropy $T_\perp/T_{||}=2,5,10,100$. The blue circles are the reference solutions derived from the Vlasov-Maxwell model, the red lines are the solution of the CJ model, the blue lines are for the neutral Vlasov model. }\label{fig:CJ_QN_ion}
 \end{center}
\end{figure}
As it was done in previous sections, the case of the electron Weibel instability is studied by adopting a cold-fluid closure for ion kinetics. Thus, upon setting ${T_\perp^{(i)}}={T_\|^{(i)}}$ and $W(\omega/{kv_{i\|}})\simeq0$, equations \eqref{disprel-CJ} and \eqref{disprel-NV} become
\begin{equation}\label{disprel-CJe}
1+\frac{T_\perp^{(e)}}{T_\|^{(e)}}W\!\left(\frac\omega{kv_{e\|}}\right)=-k^2\delta_e^2
\end{equation}
and
\begin{equation}\label{disprel-NVe}
1+\frac{T_\perp^{(e)}}{T_\|^{(e)}}\,W\!\left(\frac\omega{kv_{e\|}}\right)=-k^2\delta_e^2-Z\bmu ,
\end{equation}
respectively. We notice how the electron kinetics completely decouples from the ions in \eqref{disprel-CJe}, while a minor coupling persists in \eqref{disprel-NVe}. Similarly, the case of the ion Weibel instability can now be studied by adopting a cold-fluid closure for electron kinetics (not available in previous Sections, which instead adopted an electron warm-fluid closure for the ion Weibel instability). In this case, setting ${T_\perp^{(e)}}={T_\|^{(e)}}$ and $W(\omega/{kv_{e\|}})\simeq0$ in \eqref{disprel-CJ} and \eqref{disprel-NV} leads to
\begin{equation}\label{disprel-CJi}
Z\bmu \frac{T_\perp^{(i)}}{T_\|^{(i)}}W\!\left(\frac\omega{kv_{i\|}}\right)=-1-k^2\delta_e^2.
\end{equation}
and 
\begin{equation}\label{disprel-NVi}
Z\bmu \frac{T_\perp^{(i)}}{T_\|^{(i)}}W\!\left(\frac\omega{kv_{i\|}}\right)=-1-k^2\delta_e^2-Z\bmu \frac{T_\perp^{(i)}}{T_\|^{(i)}},
\end{equation}
respectively. We notice that certain differences between the two models may be appreciated in this case only for unusual anisotropy values of the order ${T_\perp^{(i)}}/{T_\|^{(i)}} \simeq\bmu^{-1}$.
\begin{figure}[htb!]
\begin{center}
 \noindent\includegraphics[width=0.8\textwidth]{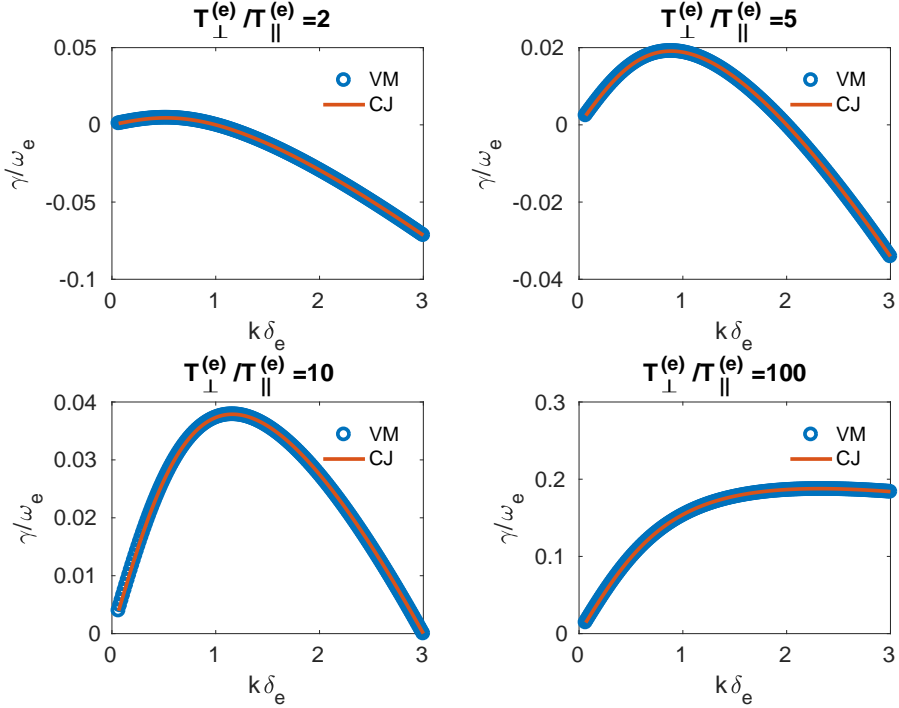}
 \caption{\footnotesize Growth rate for the electron Weibel instability for four values of temperature anisotropy $T_\perp/T_{||}=2,5,10,100$. The blue circles are the reference solutions derived from the Vlasov-Maxwell model, while the red lines are the solution of the CJ model. The neutral Vlasov solution coincides with both the VM and CJ solutions.}\label{fig:CJ_electron}
 \end{center}
\end{figure}

The dispersion relations \eqref{disprel-CJi}-\eqref{disprel-NVi} and \eqref{disprel-CJe}-\eqref{disprel-NVe} for the ion and electron Weibel instability derived from the CJ and the neutral Vlasov model are presented in Figures \ref{fig:CJ_QN_ion} and \ref{fig:CJ_electron}, respectively. The results for the electron instability are indistinguishable from the VM solutions { (obtained upon suitably specializing \eqref{equil-kin-MV} for the case of electron and ion Weibel instabilities). Neutral Vlasov solutions are not shown as they also overlap with the VM solutions.} For what concerns the ion instability a very small discrepancy can be noticed between the VM and the CJ, with the latter typically overestimating the growth/damping rates. However, the values are very small.

\vspace{-.2cm}

\section{Conclusions and perspectives}
The development and study of reduced models is a central theme in plasma physics, where the large separation of time and space scales between different species often makes computationally infeasible to tackle the first-principle dynamics (see, e.g. \cite{camporeale16}).
In this paper, { we have addressed the problem of using reduced forms of  Ohm's law to couple electrons and ions species}. We have studied the validity of different approximation schemes by studying the linear dispersion relation for both electron and ion Weibel instabilities. In a sense, this is the simplest, yet not trivial, electromagnetic instability that one would like to be able to recover in an unmagnetized plasma. The Weibel instability has important physical implications for magnetic field generation in astrophysical and cosmological scenarios \cite{fonseca,schlickeiser}.

In Section \ref{sec:HW}, we have studied a moment model, initially introduced by Hesse and Winske and later developed further by Kuztnetsova. This model was then extended to a fully  kinetic  theory that neglects electron inertia, in Section \ref{sec:kin-HW}. Also, Section \ref{sec:CJ} discussed a quasi-neutral kinetic model introduced by Cheng and Johnson, who  neglected terms of the order of the mass ratio.
Furthermore for each one of the above mentioned model, we have studied similar variants introduced through variational methods, where the approximations are introduced at the level of Hamilton's principle. In particular, the 
quasi-neutral Vlasov model (simply named \emph{neutral Vlasov}) was introduced by the authors in \cite{TrCa}.

Among the available reduced models, we have shown that only the quasi-neutral models { (with electron inertia)} are able to reproduce correctly the dispersion relation for the Weibel instability for both cases of ion and electron temperature anisotropy, { thereby} highlighting the importance of a kinetic derivation of the electron pressure tensor. 
The most important difference between the CJ and the neutral Vlasov models is that the latter preserves { Amp\`ere's current balance}. This ensures the equality, at all times, between the mean velocity calculated through the first moment of the distribution function, and the same quantity calculated through Ampere's law (\ref{neutrality}). In turn, this also guarantees charge conservation and thus { quasi-}neutrality.

\section*{Acknowledgements.} This paper belongs to the special issue for the conference Vlasovia 2016, held in Copanello (Italy); we thank all the participants interested in this work for inspiring discussions. Also, we are grateful to Joshua W. Burby, Emanuele Cazzola, Maria Elena Innocenti, Giovanni Lapenta, Giovanni Manfredi, Philip J. Morrison, and Francesco Pegoraro for stimulating conversations on this and related topics. C.T. acknowledges financial support by the Leverhulme Trust Research Project Grant No. 2014-112, and by the London Mathematical Society Grant No. 31633 (Applied Geometric Mechanics Network).

\clearpage
\newpage
%%%%%%%%%%%%%%%%%%%%%%%%%%%%%%%%%%%%%%%%%%%%%%%%

\newpage
\clearpage
\appendix

\section{Appendix}

\subsection{The dispersion relation for moment models\label{app:1}}

In this Appendix, we derive the dispersion relations \eqref{dispHW} and \eqref{dispHW-m} for the moment models treated in Section \ref{sec:moments}.

First, we decompose all quantities as $X=X_0+X_1$ (where the subscripts 0 and 1 denote the equilibrium configuration and its perturbation, respectively). Then, we linearize the ion Vlasov equation to find
\[
\frac{\partial f_1}{\partial t}+\bv\cdot\frac{\partial f_1}{\partial \bq}
+\bigg[\frac{e}{m_e}\Big(\frac{\partial\bA_1}{\partial t}+\nabla\varphi_1-\bv\times\bB_1\Big)
\bigg]\cdot\frac{\partial f_0}{\partial \bv}=0\,,
\]
where we have dropped the subscript $i$ for convenience of notation. Upon applying the method of characteristics \cite{KrTr}, we write $X_1=\widetilde{X}_1e^{kz-\omega t}$ and find
\begin{align}\nonumber
\tilde{f}_1=&-i\frac{Ze}{m_i}\int_{-\infty}^0\left(\omega\tA-\widetilde{\varphi}_1\bk+\bv\times\bk\times\tA\right)\cdot\frac{\partial f_0}{\partial \bv} \,e^{i(kv_z-\omega)\tau}\,\de\tau
\\
=&
\frac{Ze}{m_i}\bigg[{\tA}+\frac{\widetilde{\varphi}_1-\bv\cdot\tA}{kv_z-\omega}\bk
\bigg]\cdot\frac{\partial f_0}{\partial \bv}
\,.
\label{kin-ion-sol}
\end{align}
At this point, we denote by $\bK_i=\int\!\bv{f}_i\,\de\bv$ the momentum of the ion mean flow and we compute its planar projection (by dropping the subscript $i$) as
\begin{align}\nonumber
\widetilde{\mathbf{K}}_{1\perp}=&\frac{Ze}{m_i}\int\!
\bv_\perp\left(\tA_\perp\cdot\frac{\partial f_0}{\partial \bv_\perp} \right)
-\frac{Ze}{m_i}\int k\bv_\perp\frac{\bv_\perp\cdot\tA_\perp}{kv_z-\omega}\frac{\partial f_0}{\partial v_z}
\\
=&-\frac{e}{m_i}n_0\tA_\perp-\frac{Ze}{m_i}\tA_\perp\int\frac12\frac{kv_\perp^2}{kv_z-\omega}\frac{\partial f_0}{\partial v_z}
\\
=&-\frac{e}{m_i}n_0\left[1+\frac{T_\perp^{(i)}}{T_\|^{(i)}}\,W\!\left(\frac{\omega}{kv_{i\|}}\right)\right]\tA_\perp\,,
\end{align}
where we have used the fact that $f_0=f_0(v_\perp^2,v_z^2)$ (notice, $f_0$ is an even function of $v_\perp$) and so $\int\!\bv_\perp\bv_\perp f_0\,\de\bv_\perp=\left(\int (v_\perp^2/2) f_0\,\de\bv_\perp\right)\boldsymbol{1}$ (here, $\boldsymbol{1}$ denotes the identity matrix). Also, here we have introduced the superscript $(i)$ on the ion temperatures as well as the notation
\[
W(\xi)=-1-\xi\mathcal{Z}(\xi)
\,,
\]
where $\mathcal{Z}$ denotes the plasma dispersion function. In addition, here $v_{i\|}$ denotes the ion thermal velocity in the parallel direction. As a subsequent step, we linearize Amp\`ere's law $Ze\bK_i-en_e\bV_{\!e}=\mu_0^{-1}\nabla\times\bB$ to obtain
\begin{equation}\label{amp-lin}
\widetilde{\mathbf{K}}_{1\perp}=\frac{n_0}Z\left(\!\tV_\perp+\frac{e}{m_e}k^2\delta^2_e\tA_\perp\!\right)
\end{equation}
so that eventually
\[
-\frac{e}{m_i}n_0\left[1+\frac{T_\perp^{(i)}}{T_\|^{(i)}}\,W\!\left(\frac{\omega}{kv_{i\|}}\right)\right]\tA_\perp=\frac{n_0}Z\left(\!\tV_\perp+\frac{e}{m_e}k^2\delta^2_e\tA_\perp\!\right)
\]
and thus
\begin{equation}\label{auxeq1}
\tV_\perp=-\left\{\frac{e}{m_e}k^2\delta^2_e+\frac{Ze}{m_i}\left[1+\frac{T_\perp^{(i)}}{T_\|^{(i)}}\,W\!\left(\frac{\omega}{kv_{i\|}}\right)\right]\right\}\tA_\perp\,.
\end{equation}

At this point, we linearize the electron pressure equation to obtain
\begin{multline*}
\omega\tP-\tV(\bk\cdot\Bbb{P}_0)
-(\bk\cdot\tV)\Bbb{P}_0
-(\bk\cdot\Bbb{P}_0)\tV
\\
-\alpha\Big({\Bbb{P}}_0\times(\bk\times\tV)
-(\bk\times\tV)\times{\Bbb{P}_0}\Big)
+i\frac{e}{m_e}\Big({\Bbb{P}}_0\times\tB-\tB\times{\Bbb{P}}_0\Big)
=0
\,.
\end{multline*}
Notice that we have included the Coriolis force terms for completeness; when $\alpha=0$, the equation above returns the HW model, while $\alpha=1$ retains the Coriolis force consistently. Then, we take the dot product of the equation above with $\bk$ (strictly speaking, we contract the pressure tensor equation with the vector $\bk$). To this purpose, we compute
\begin{align*}
(\bk\cdot{\Bbb{P}}_0)\times(\bk\times\tV)=&\ 
(\bk\tV:\Bbb{P}_0)\bk-(\bk\bk:\Bbb{P}_0)\tV
\\
\bk\cdot\big((\bk\times\tV)\times{\Bbb{P}_0}\big)
=&\, 
\bk\cdot\!\int\!\!\big((\bk\times\tV)\times\bc\big)\bc\,f_0(\bc)\,\de^3\bc
\\
=&\, 
-\bk\cdot\!\int\!\!\big((\bc\cdot\tV)\bk-(\bk\cdot\bc)\tV\big)\bc\,f_0(\bc)\,\de^3\bc
\\
=&\,-k^2(\tV\cdot\Bbb{P}_0)+(\bk\cdot\tV)\Bbb{P}_0\bk
\,,
\end{align*}
so that eventually
\[
\omega\tP\bk
=(1-\alpha)p_\parallel(\bk\cdot\tV)\bk+
\big(\alpha p_\perp+(1-\alpha)p_\parallel\big)k^2\tV+i\frac{e}{m_e}
(p_\perp-p_\parallel){\bk}\times\tB
\,.
\]
Taking the planar projection yields
\begin{equation}\label{auxeq2}
\omega(\tP\bk)_\perp
=
\big(\alpha p_\perp+(1-\alpha)p_\parallel\big)k^2\tV_\perp+
\frac{e}{m_e}(p_\perp-p_\parallel){k}^2\tA_\perp
\end{equation}

Now, we linearize Ohm's law \eqref{ohmslaw1} to write $
en_0(\widetilde{\varphi}_1\bk-\omega\tA)=\tP\bk
%-\beta m_en_e\omega\tV
$.
%Here, we have inserted the flag $\beta$ so that $\beta=1$ returns the general Ohm's law \eqref{ohmslaw1}, while the HW model is recovered by $\beta=0$. 
Taking the planar component of the latter equation yields
\begin{equation}\label{lin-ohm}
-en_0\omega^2\tA_\perp=\omega(\tP\bk)_\perp
%-\beta m_en_e\omega^2\tV_\perp
\end{equation}
and by inserting the equations \eqref{auxeq1} and \eqref{auxeq2} we obtain
\begin{multline*}
-en_0\omega^2\tA_\perp=\frac{e}{m_e}(p_\perp-p_\parallel){k}^2\tA_\perp
\\
\qquad\qquad
-\big(\alpha p_\perp+(1-\alpha)p_\parallel\big)k^2\left\{\frac{Ze}{m_i}\left[1+\frac{T_\perp^{(i)}}{T_\|^{(i)}}\,W\!\left(\frac{\omega}{kv_{i\|}}\right)\right]+\frac{e}{m_e}k^2\delta^2_e\right\}\tA_\perp\qquad\qquad\qquad\qquad
%-\beta m_en_e\omega^2\left\{\frac{Ze}{m_i}\left[1+\frac{T_\perp^{(i)}}{T_\|^{(i)}}\,W\!\left(\frac{\omega}{kv_{i\|}}\right)\right]-\frac{e}{m_e}k^2\delta^2_e\right\}\tA_\perp
.
\end{multline*}
Then, upon using the relation $m_en_0/p_\|=1/v_{e\|}$, the dispersion relation becomes
\[
%\begin{multline*}
%\left\{1
%+\beta \left[k^2\delta^2_e-\frac{Zm_e}{m_i}\left(1+\frac{T_\perp^{(i)}}{T_\|^{(i)}}\,W\!\left(\frac{\omega}{kv_{i\|}}\right)\right)\right]\!\right\}
\frac{\omega^2}{k^2 v_{e\|}^2}=
%\\
1-
\frac{T_\perp^{(e)}}{T_\parallel^{(e)}}
+
\left(\alpha \frac{T_\perp^{(e)}}{T_\parallel^{(e)}}+1-\alpha\right)\left\{\frac{Zm_e}{m_i}\left[1+\frac{T_\perp^{(i)}}{T_\|^{(i)}}\,W\!\left(\frac{\omega}{kv_{i\|}}\right)\right]+k^2\delta^2_e\right\}
\,.
%\end{multline*}
\]
Then, the dispersion relation \eqref{dispHW} for the HW model in the case of the Weibel instability is given by $\alpha=0$
\[
\frac{\omega^2}{k^2 v_{e\|}^2}=
1-
\frac{T_\perp^{(e)}}{T_\parallel^{(e)}}
+
\left\{\frac{Zm_e}{m_i}\left[1+\frac{T_\perp^{(i)}}{T_\|^{(i)}}\,W\!\left(\frac{\omega}{kv_{i\|}}\right)\right]+k^2\delta^2_e\right\}
\,,
\]
while retaining Coriolis effects (by setting $\alpha=1$) leads to \eqref{dispHW-m}{ , that is}
\[
\frac{\omega^2}{k^2 v_{e\|}^2}=
1-
\frac{T_\perp^{(e)}}{T_\parallel^{(e)}}\left\{
1-k^2\delta^2_e-
\frac{Zm_e}{m_i}\left[1+\frac{T_\perp^{(i)}}{T_\|^{(i)}}\,W\!\left(\frac{\omega}{kv_{i\|}}\right)\right]
\right\}
\,.
\]

\subsection{The dispersion relation for reduced kinetic models\label{app:2}}
This Appendix presents the dispersion relations \eqref{dispHWkin} and \eqref{dispHWkin-m} for the reduced models in Sections \ref{sec:kin-HW} and \ref{sec:kin-Tronci}, in the case of the Weibel instability. The ion kinetic equation \eqref{ion-f} was already linearized around the equlibrium
\begin{equation}%\label{equil2}
\bE_0=\bB_0=\bV_{\!e\,0}=\bV_{\!i\,0}=0
\,,\qquad
f_0=f_0(v_\perp^2,v_z^2)
\end{equation}
 in Appendix \ref{app:1}, thereby leading to \eqref{kin-ion-sol}. In addition, in Appendix \ref{app:1}, linearising Amp\`ere's law led to \eqref{amp-lin} and to \eqref{auxeq1}. At this point, we need to linearize electron kinetics around the equilibrium \eqref{equil-kin}. To this purpose, we consider the following equation
 \[
 \frac{\partial \mathsf{f}_e}{\partial t}+\big(\bc+\bV_e\big)\cdot\frac{\partial \mathsf{f}_e}{\partial \bq}
-\bigg[\bc\cdot\nabla\bV_e+\alpha\bc\times\nabla\times\bV_e-\frac{e}{m_e}\Big(\frac{\partial\bA}{\partial t}+\nabla\varphi-(\bc+\bV_e)\times\bB\Big)
\bigg]\cdot\frac{\partial \mathsf{f}_e}{\partial \bc}=0
\,,
 \]
where $\alpha=0,1$ is a flag variable so that $\alpha=0$ corresponds to the kinetic HW system in Section \ref{sec:kin-HW}, while $\alpha=1$ corresponds to its variational variant in Section \ref{sec:kin-Tronci}. Upon linearising around \eqref{equil-kin}, we obtain
\[
\frac{\partial \mathsf{f}_1}{\partial t}+\bc\cdot\frac{\partial \mathsf{f}_1}{\partial \bq}
-\bigg[\frac{q_e}{m_e}\Big(\frac{\partial\bA_1}{\partial t}+\nabla\varphi_1-\bc\times\bB_1\Big)
+\bc\cdot\nabla\bV_1+\alpha\bc\times\nabla\times\bV_1\bigg]\cdot\frac{\partial \mathsf{f}_0}{\partial \bc}=0
\,,
\]
where we have used the same notation as in Appendix \ref{app:1}. Again, upon using the method of characteristics and by Fourier-transforming, we have
\begin{align*}
\widetilde{\mathsf{f}}_1=&\int_{-\infty}^0\bigg[\zeta_e\Big({i\omega\tA}-i\widetilde{\varphi}_1\bk+\bc\times\tB\Big)
+ikc_z\tV+i\alpha\bc\times\bk\times\bV_1\bigg]\cdot\frac{\partial \mathsf{f}_0}{\partial \bc}e^{i(kc_z-\omega)\tau}\de\tau
\\
=&\ 
i\!\int_{-\infty}^0\bigg\{\zeta_e{(\omega-kc_z)\tA}-\big[\zeta_e\widetilde{\varphi}_1-\bc\cdot(\zeta_e\tA+\alpha\tV)\big]\bk
\\&\hspace{8.4cm}
+(1-\alpha)kc_z\tV\bigg\}\cdot\frac{\partial \mathsf{f}_0}{\partial \bc}e^{i(kc_z-\omega)\tau}\de\tau
\\
=&\ 
-\bigg[\zeta_e{\tA}+\frac{\zeta_e\widetilde{\varphi}_1-\bc\cdot(\zeta_e\tA+\alpha\tV)}{kc_z-\omega}\bk
+\frac{
(\alpha-1)kc_z}{kc_z-\omega}\tV\bigg]\cdot\frac{\partial \mathsf{f}_0}{\partial \bc}
\,,
\end{align*}
where we have introduced the notation $\zeta_e=e/m_e$. Therefore, the planar components of the pressure force term are
\begin{align*}
m_e^{-1}(\tP\bk)_\perp=-\int\!kc_z\left\{\bigg[\zeta_e{\tA}+\frac{\zeta_e\widetilde{\varphi}_1-\bc\cdot(\zeta_e\tA+\alpha\tV)}{kc_z-\omega}\bk
+\frac{
(\alpha-1)kc_z}{kc_z-\omega}\tV\bigg]\cdot\frac{\partial \mathsf{f}_0}{\partial \bc}\right\}\bc_\perp\,\de^3\bc
\end{align*}
Then, we verify that
\[
\int\!\left(kc_z{\tA}\cdot\frac{\partial \mathsf{f}_0}{\partial \bc}\right)\bc_\perp\,\de^3\bc
=k\int\!\widetilde{\varphi}_1\frac{kc_z}{kc_z-\omega}
\frac{\partial \mathsf{f}_0}{\partial c_z}\bc_\perp\,\de^3\bc
=
k^2\,\widetilde{\!A}_{1\,z\!}\int\!\frac{c_z^2}{kc_z-\omega}
\frac{\partial \mathsf{f}_0}{\partial c_z}\,\bc_\perp\,\de^3\bc=0
\]
and also
\[
\int\!kc_z\bigg[\frac{kc_z(\alpha\widetilde{V}_{1z}+\zeta_e\,\widetilde{A}_{1z})}{kc_z-\omega}
-\frac{(\alpha-1)kc_z}{kc_z-\omega}\widetilde{V}_{1z}\bigg]\frac{\partial \mathsf{f}_0}{\partial c_z}\,\bc_\perp\,\de^3\bc=0
\,.
\]
With this in mind, and by recalling $\int\!\bv_\perp\bv_\perp \mathsf{f}_0\,\de\bv_\perp=\left(\int (v_\perp^2/2) \mathsf{f}_0\,\de\bv_\perp\right)\boldsymbol{1}$, we compute
\begin{align}\nonumber
m_e^{-1}(\tP\bk)_\perp=&\int\!kc_z\left\{\bigg[\frac{k\bc_\perp\cdot(\alpha\tV_\perp+\zeta_e\,\widetilde{\!\bA}_{1\,\perp})}{kc_z-\omega}\frac{\partial \mathsf{f}_0}{\partial c_z}
-\frac{
(\alpha-1)kc_z}{kc_z-\omega}\tV_\perp\cdot\frac{\partial \mathsf{f}_0}{\partial \bc_\perp}\bigg]\right\}\bc_\perp
\\
=&
\int\!kc_z\left[\frac12\frac{{kc_\perp^2}{\partial \mathsf{f}_0}/{\partial c_z}}{kc_z-\omega}(\alpha\tV_\perp+\zeta_e\,\widetilde{\!\bA}_{1\,\perp})
+\frac{(\alpha-1)kc_z}{kc_z-\omega}\mathsf{f}_0\tV_\perp \right]
\nonumber
\\
=&\ 
n_0\omega\frac{T_\perp^{(e)}}{T_\|^{(e)}} (\alpha\tV_\perp+\zeta_e\,\widetilde{\!\bA}_{1\,\perp})\,W\!\left(\frac{\omega}{kv_{e\parallel}}\right)
+(\alpha-1)\omega\!\left[n_0+\omega\int\!\frac{\mathsf{f}_0}{kc_z-\omega}\right]\tV_\perp 
\nonumber
\\
=&\ 
n_0\omega\left[\frac{T_{\perp}^{(e)}}{T_{\|}^{(e)}} (\alpha\tV_\perp+\zeta_e\,\widetilde{\!\bA}_{1\,\perp})+(1-\alpha)\tV_\perp\right]W\!\left(\frac{\omega}{kv_{e\parallel}}\right),
\label{pressure-comp}
\end{align}
Now, the planar components of Ohm's law (as in \eqref{lin-ohm}) yield
\[
-\frac{e}{m_e}\left[1+\frac{T_{\perp}^{(e)}}{T_{\|}^{(e)}}\,W\!\left(\frac{\omega}{kv_{e\parallel}}\right)\right]\tA_\perp=\left[1-\alpha\left(1-\frac{T_{\perp}^{(e)}}{T_{\|}^{(e)}}\right) \right]W\!\left(\frac{\omega}{kv_{e\parallel}}\right)\tV_\perp
\,,
\]
so that, upon recalling \eqref{auxeq1},
\[
1+\frac{T_{\perp}^{(e)}}{T_{\|}^{(e)}}\,W\!\left(\frac{\omega}{kv_{e\parallel}}\right)
=\left[1-\alpha\left(1-\frac{T_{\perp}^{(e)}}{T_{\|}^{(e)}}\right) \right]\left\{k^2\delta^2_e+{Z\bmu}\left[1+\frac{T_\perp^{(i)}}{T_\|^{(i)}}\,W\!\left(\frac{\omega}{kv_{i\|}}\right)\right]\right\}W\!\left(\frac{\omega}{kv_{e\parallel}}\right).
\]
Then, the dispersion relation \eqref{dispHWkin} for the HW { kinetic} model in the case of the Weibel instability is given by $\alpha=0$
\[
1+\frac{T_{\perp}^{(e)}}{T_{\|}^{(e)}}\,W\!\left(\frac{\omega}{kv_{e\parallel}}\right)
=\left\{k^2\delta^2_e+{Z\bmu}\left[1+\frac{T_\perp^{(i)}}{T_\|^{(i)}}\,W\!\left(\frac{\omega}{kv_{i\|}}\right)\right]\right\}W\!\left(\frac{\omega}{kv_{e\parallel}}\right)
\,,
\]
while retaining Coriolis effects (by setting $\alpha=1$) leads to \eqref{dispHWkin-m}{ , that is}
\[
\frac{T_{\|}^{(e)}}{T_{\perp}^{(e)}}+W\!\left(\frac{\omega}{kv_{e\parallel}}\right)
=\left\{k^2\delta^2_e+{Z\bmu}\left[1+\frac{T_\perp^{(i)}}{T_\|^{(i)}}\,W\!\left(\frac{\omega}{kv_{i\|}}\right)\right]\right\}W\!\left(\frac{\omega}{kv_{e\parallel}}\right).
\]

\bigskip

\subsection{Dispersion relation for the Cheng-Johnson model}\label{app:3}
This appendix presents the dispersion relation \eqref{disprel-CJ} that arises by linearizing the Cheng-Johnson model around the equilibrium \eqref{equil-kin-CJ}. In this case, linearizing the form \eqref{ohmCJ} of Ohm's law leads to
\begin{equation*}
\frac{q_e^2}{m_e}n_0\bE_1+\mu_0^{-1}\nabla\times\nabla\times\bE_1={q_e}\frac1{m_e}\nabla\cdot{\Bbb{P}}_{e\,1}
+{q_i}\frac1{m_i}\nabla\cdot{\Bbb{P}}_{i\,1}
\end{equation*}
and taking the planar components {after Fourier transforming} leads to
\[
\omega(1+k^2\delta^2)\tA_\perp=\frac{1}{en_0}\left[(Z\bmu\,\tP^{(i)}-\tP^{(e)})\bk\right]_\perp
\,.
\]
On the other hand, by adapting the result \eqref{pressure-comp} to the present case, we have
\[
m_s^{-1}(\tP^{(s)}\bk)_\perp=\pm\omega\frac{en_0}{m_s}\frac{T_\perp^{(s)}}{T_\|^{(s)}}W\!\left(\frac\omega{kv_{s\|}}\right)\tA_\perp
\]
(where the plus is used when $s=e$ and the minus when $s=i$)
and therefore we obtain \eqref{disprel-CJ} {in the form}
\[
1+k^2\delta_e^2=-\frac{T_\perp^{(e)}}{T_\|^{(e)}}W\!\left(\frac\omega{kv_{e\|}}\right)-Z\bmu \frac{T_\perp^{(i)}}{T_\|^{(i)}}W\!\left(\frac\omega{kv_{i\|}}\right).
\]

\end{document}